\documentclass[twocolumn,showpacs,preprintnumbers,amsmath,amssymb]{revtex4}
\usepackage{graphicx}

\begin{document}
\title{\large{Ferroelectric Phase Transitions in Ultra-thin Films of BaTiO$_3$
}}
\author{Jaita Paul$^1$, Takeshi Nishimatsu,$^2$$^,$$^3$$^,$$^4$ Y. Kawazoe$^2$ and 
Umesh V. Waghmare$^1$}
\affiliation{$^1$Theoretical Sciences Unit\\
Jawaharlal Nehru Centre for Advanced Scientific Research\\ Jakkur PO
Bangalore 560 064 India
}
\affiliation{
$^2$Institute for Materials Research\\ Tohoku University\\
Sendai, 980-8577, Japan}
\affiliation{$^3$Department of Physics and Astronomy, Rutgers University, 136 
Frelinghuysen Road, Piscataway, NJ 08544-8019\\
$^4$Japan Society for the Promotion of Science (JSPS) Post Doctoral Fellow for 
Research Abroad} 

\begin{abstract}
We present molecular dynamics simulations of a realistic model of an ultrathin
film of BaTiO$_3$ sandwiched between short-circuited electrodes to determine and understand
effects of film thickness, epitaxial strain and the nature of electrodes on
its ferroelectric phase transitions as a function of temperature. We determine 
a full epitaxial strain-temperature phase diagram in the presence of
perfect electrodes. Even with the vanishing depolarization field, we find
that ferroelectric phase transitions to states with in-plane and out-of-plane 
components of polarization exhibit dependence on thickness; it arises from the
interactions of local dipoles with their electrostatic images in the presence of electrodes. 
Secondly, in the presence of relatively bad metal electrodes which only partly
compensate the surface charges and depolarization field, a qualitatively
different phase with stripe-like domains is stabilized at low temperature.

\end{abstract}
\pacs{}
\maketitle


Ferroelectric (FE) materials are vital to the technologies based on
micro and nano electromechanical systems and certain random access memories.
The fact that ferroelectricity in finite objects is very sensitive to 
mechanical and electrical boundary conditions is of great relevance 
to nano-scale devices based on ferroelectric materials \cite{scott, Ahn}. 
These boundary conditions in the case of a thin film are determined by 
the properties of its interfaces with substrate on which it is  
grown and the electrodes. While understanding the suppression of ferroelectricity
in perovskite thin films has been a fundamental issue \cite{fong},
a clever choice of an electrode and a substrate can be used effectively
in designing nano-scale ferroelectric film based structures with 
desired properties. For example, a high-T$_c$ lead-free ferroelectric 
was developed by growing BaTiO$_3$ films on an appropriate substrate \cite{choi}.

FE materials are spontaneously polarized below a certain 
temperature and the direction of this spontaneous polarization 
can be switched by an external electric field \cite{lines}. When a FE film
is polarized in a direction perpendicular to its plane, the bound charges at 
the surface of the film give rise to depolarization field that suppresses 
the spontaneous polarization. Free carriers in the interfacing electrodes partially
compensate these surface charges, and aid in maintaining the ferroelectricity
down to 6 unit cell thicknesses \cite{ghosez}. While the suppression of the 
out-of-plane polarization has been investigated fairly well \cite{ghosez,fong,choi}
and can be understood with relative ease, effects of electrodes on the in-plane 
polarization and associated ferroelectricity have not been explored yet; their
origin is expected to be fundamentally different.
Since the switching of polarization necessary for memory applications
occurs by rotation \cite{fu-cohen} through the states with in-plane polarization, 
understanding properties of these states is very important to applications 
of FE materials. 

BaTiO$_3$, a standard example of FE perovskite \cite{lines}, exhibits a 
sequence
of phase transitions from cubic paraelectric to tetragonal (C$<$-$>$T) to 
orthorhombic (T$<$-$>$O) to 
rhombohedral (O$<$-$>$R) ferroelectric phases characterized by polarization 
ordering along
(001), (110) and (111) directions respectively.
It is an ideal case for exploration
of effects of electrodes, epitaxial strain and film thickness on in-plane
ferroelectricity. Phenomenological Landau theory has been used to map
a temperature-misfit strain phase diagram of BaTiO$_3$ and PbTiO$_3$ films of 
thickness $>$ 50 nm grown on cubic substrate, with short circuit boundary 
conditions \cite{pertsev}. Di\'{e}guez et al., \cite{vand} mapped the 
equilibrium BaTiO$_3$
structure at zero temperature as a function of epitaxial strain and simulated
corresponding ferroelectric transitions in the bulk using the effective-Hamiltonian 
approach \cite{zhong, rabe, umesh}. Information about the strain-temperature 
phase diagram is very useful in the design of ferroelectric nano-structures.

In this letter, we use a first-principles model Hamiltonian of BaTiO$_3$ 
of Ref. \onlinecite{zhong} along with a simple electrostatic model of 
electrodes \cite{dawber}
in classical molecular dynamics simulations and determine (a) thickness dependence of
ferroelectric transitions in BaTiO$_3$ films sandwiched between limiting cases of
electrodes (perfect vs. imperfect), (b) a complete epitaxial strain-temperature phase
diagram in the case of perfect electrodes and (c) the nature of low-temperature phase
and ferroelectricity when the electrodes are not so perfect. We point out that interactions
with electrostatic images of the in-plane dipoles interacting are responsible for
size dependent ferroelectricity, even in the absence of depolarization field.

\begin{figure}[htbp]
\centering
\includegraphics[scale=0.42]{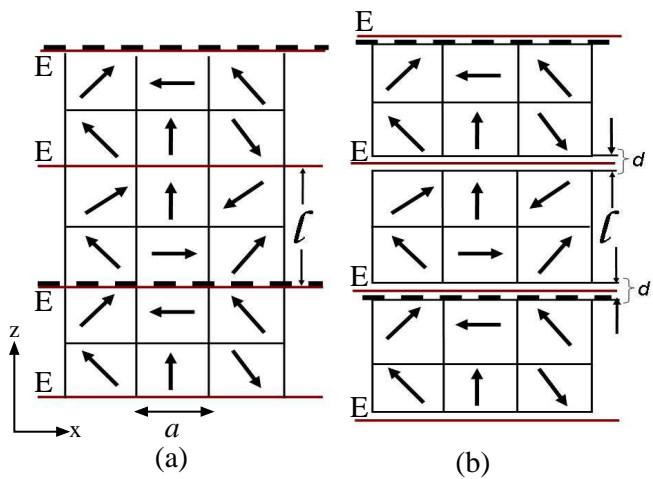}
\caption{Schematic representation of a FE thin film of thickness $l$ unit cells (here
$l$=2) sandwiched between (a) perfect and (b) imperfect electrodes. Horizontal thick
lines marked as ``E" represent the electrostatic mirrors used to model electrodes. 
They are distance $\frac{d}{2}$ away from the FE film surface ($d$=0 in (a), $d$=1 in
(b)). Each arrow represents a local dipole within a unit cell ($a$=3.94 \AA \space) of
the FE crystal. Thick dashed lines indicate boundaries of the periodic box of the 
system used in simulations.}
\end{figure}


We used an effective lattice dynamical (phonon) Hamiltonian of BaTiO$_3$ expressed
in terms of localized atomic displacement basis spanned by two classical 
3-dimensional vector degrees of freedom (DoF) per unit cell: one representing
the unstable polar optical branches of the cubic structure and another 
representing the branches of acoustic phonons. 
This amounts to approximately
integrating out the high energy phonons, reducing DoFs from 15 to 6 per unit cell.
The average value of the former
yields polarization, where as the latter describes local strains in the crystal.
In addition, six components of the homogeneous strain tensor are included in
the DoFs. This Hamiltonian includes harmonic and anharmonic interactions of
phonons, coupling between polar and strain DoFs, and elastic contribution to
energy. We use the parameters in effective Hamiltonian as 
determined earlier from first-principles local density functional theory
calculations \cite{zhong, rabe}. Due to underestimation of lattice constant in 
this scheme,
a negative pressure of -5 GPa is used in all simulations for comparison with experiment.
An effective mass is attached to each polar DoF, based on the eigenvector of
the unstable mode at $\Gamma-$point and ionic masses.

To model a FE thin film sandwiched between two electrodes, we have treated electrodes
as perfect electrostatic mirrors located at a distance $\frac{d}{2}$ away from the 
surface
of the film (see Fig. 1).  This ``gap" can also be thought of as a dead layer 
separating 
electrodes from the FE film. The length-scale $d$ determines how effective the 
electrodes are in compensating the depolarization field \cite{dawber}:
\begin{eqnarray}
\centering
\centerline{$E_{d} = -4\pi\frac{d}{l+d}P_{z}$\^z}
\end{eqnarray}
where $P_{z}$ is the out-of-plane polarization in the film, $l$ being the thickness
of film (both $l$ and $d$ are in the units of number of unit cells). Here, 
we use two 
limiting values of $d$: $d=0$ corresponding to ``perfect" electrodes
and vanishing depolarization field, and $d=1$ corresponding to ``bad" electrodes
on the BaTiO$_3$ thin films. It is clear that an electrostatic image of an electric 
dipole
in the FE film is a dipole with same magnitude inside the electrode. However, the 
direction of the image dipole is the same (opposite) as that of the dipole in the film, if
it is in $z$ ($x$ or $y$) direction. In this work, we consider identical electrodes 
on the two sides of the FE film, which allows simulations with periodic boundary
conditions (due to infinite number of images, see Fig.1)
with a periodicity of 2($l$+$d$), i.e., treatment of electrodes amounts to 
simulating periodic system with twice the size of the film.
We simulated FE films of two types: (a) bulk-like (system $F$)
which have no epitaxial constraint and 
can be strained in the plane of the films, and (b) epitaxial films (system $EF$), 
whose in-plane strain 
is fixed by the choice of the substrate (electrode). We keep the in-plane homogeneous
strain DoF fixed during the simulation of the latter. 
As a check on internal consistency
and for comparison, we also simulated ferroelectricity in bulk 
BaTiO$_3$ (system $B$) with the same size(s) using periodic ({\emph{not mirror}}) 
boundary conditions (note that there are no electrodes in system B, unlike in
systems $F$ and $EF$).

Mixed-space molecular dynamics, used earlier in large-scale simulations of 
relaxors \cite{umesh1, umesh2} treating infinite-range dipolar interactions
in reciprocal space, is used here to determine finite temperature properties
of the effective Hamiltonian. The Nos\`{e}-Poincar\`{e} \cite{nose} thermostat
with simplectic properties implemented in the present simulations allows us
to use a relatively large timestep of 2 femtoseconds. At each temperature,
the system was thermalized with 50,000 timesteps and averaging was
performed using configurations of subsequent 150,000 timesteps (amounting to
a simulation of 0.3 nano sec). Since ferroelectric phase transitions are known
to be first-order and exhibit hysteresis, we increased (decreased) temperature
in small steps of 10 to 20 K in the heating (cooling) runs of simulations, with
smaller temperature steps near the transition.
The size of systems in our simulations is $16\times16\times l$, where $l=2,11$ 
unit cells.

\begin{figure}[htbp]
\centering
\includegraphics[scale=0.35,angle=-90]{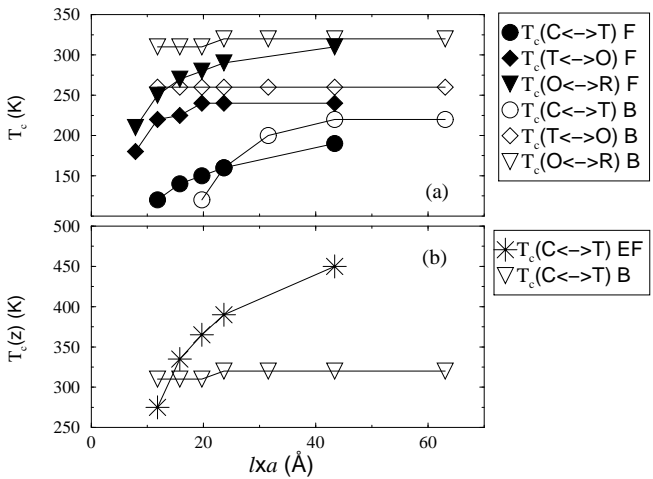}
\caption{
Dependence of transition temperatures on the thickness of films
sandwiched by perfect electrodes (a) with no and (b) with epitaxial constraint
(B: Bulk, given for comparison, F: Film without epitaxial strain, EF: Films 
with epitaxial strain): $a$ = lattice constant of BaTiO$_3$.}
\end{figure}

We first report our results for FE films sandwiched between perfect electrodes:
$d=0$.
In the limit of large thickness, ferroelectric properties of thin films with no 
epitaxial constraint (system $F$) are expected to converge to the bulk behavior.
We find for both the systems $F$ and $B$ that the transition temperatures do depend on
the thickness for polarization ordering in all the three directions of 
polarization (Fig. 2a). The dependence of $T_c$ of the system $B$ arises from the 
statistical mechanical effects of the finite size \cite{SM}, where
as the difference between the $T_c$'s of systems $B$ and $F$ reveal the nano-size 
effects of the film thickness. It is evident that the transitions to FE states
with in-plane polarization occurring at higher temperatures in the films converge to bulk
behavior when the film thickness is greater than 10 unit cell layers, where as the transition
to the state with out-of-plane polarization converges more gradually to that in the bulk.
This dependence of $T_c(z)$ on the film thickness even in the case of vanishing depolarization
field is surprising. 

A related finding from our simulations is that even the high temperature paraelectric 
phase $P=0$ of films $F$ exhibits tetragonality through uniaxial symmetry of its
dielectric constant. The origin of both the findings can be traced to the fact that
electrostatic images of the in-plane local dipoles in the presence of electrodes  
are inverted and break the horizontal reflection symmetry $\sigma_h$. Since 
these images interact with local dipoles in the film polarized in any general direction,
in a way distinct from that in the bulk,
the presence of electrodes gives rise to thickness dependence of ferroelectric transitions
in films even in the absence of depolarization field. 

We find a rather different thickness dependence of the ferroelectric transition behavior 
in epitaxial BaTiO$_3$ films sandwiched between perfect electrodes (see Fig. 2b).
In this case, the in-plane strain is clamped which suppresses the fluctuations associated
with in-plane dipole moments (FE transitions are fluctuation driven first-order phase
transitions \cite{umesh}) and hence the transition temperature for in-plane 
ordering is reduced, 
which remains almost constant at 70 K ($\pm$10 K) as a function of film
thicknesses. Unlike bulk, there is no orthorhombic phase stable at intermediate temperatures.
Epitaxial constraint naturally favors uniaxiality, resulting in a noticeable enhancement of 
the temperature of the transition (well above the bulk transition) to the state 
with out-of-plane polarization, consistent with
experimental observations \cite{choi}. Our simulations suggest a critical thickness of 3 unit 
cell layers, below which we find no ordering of $P_z$ at low temperatures.

The value of epitaxial strain has a spectacular effect on ferroelectric phase transitions
in the films (see Fig. 3 for results in the presence of perfect electrodes $d=0$). 
At compressive epitaxial strains ($\epsilon_{xx} < -0.005$), BaTiO$_3$ films exhibit
a single transition to tetragonal phase, with increasing $T_c$ with the magnitude of strain.
For $-0.005 < \epsilon_{xx} < 0.03$, they exhibit two transitions from paraelectric to tetragonal FE
and then distorted rhombohedral FE phases ($R^{'}$) as the temperature is lowered. There is a crossover point
at $\epsilon_{xx}=0.012$, below (above) which the low temperature phase has $P_z>P_x=P_y$ ($P_z<P_x=P_y$);
at this point, there is only one second order transition from cubic to rhombohedral FE phase, which
does not seem to depend on the film thickness. For $\epsilon_{xx} > 0.03$, there is a single phase
transition to orthorhombic FE phase whose $T_c$ increases with the magnitude of strain.
Through comparison with the work of Di\'{e}guez et al \cite{vand}, we show (a) the presence of 
perfect electrodes results in a shift in the temperature-strain phase diagram to the right along
strain axis by roughly $\Delta \epsilon_{xx} \sim 0.012 $, and (b) and there is an overall 
reduction in the transition temperatures by about 70 K partly due to the dependence on film
thickness. General features of our phase diagram are closer to that determined by Di\'{e}guez et al \cite{vand}
than the one of Pertsev et al \cite{pertsev}.
We note that the low-temperature aspects of this phase diagram are
consistent with properties of BaTiO$_3$/SrTiO$_3$ superlattice as a function of
epitaxial strain \cite{leejun}.

\begin{figure}[htbp]
\centering
\includegraphics[scale=0.35,angle=-90]{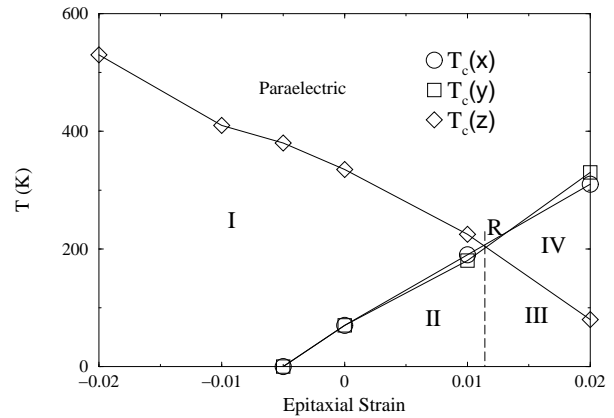}
\caption{
Temperature-epitaxial strain phase diagram of a film (of 4 unit cell thickness) 
sandwiched between perfect electrodes ($d=0$).
Region I is tetragonal, II and III are the $R^{'}$ phase (see text and [Ref.8]), 
and IV is orthorhombic. Regions II (P$_x$ (= P$_y$) $<$ P$_z$) and
III (P$_x$ (= P$_y$) $>$ P$_z$) are separated by a rhombohedral phase R. }
\end{figure}

\begin{figure}
\centering
\includegraphics[scale=0.3,angle=-90]{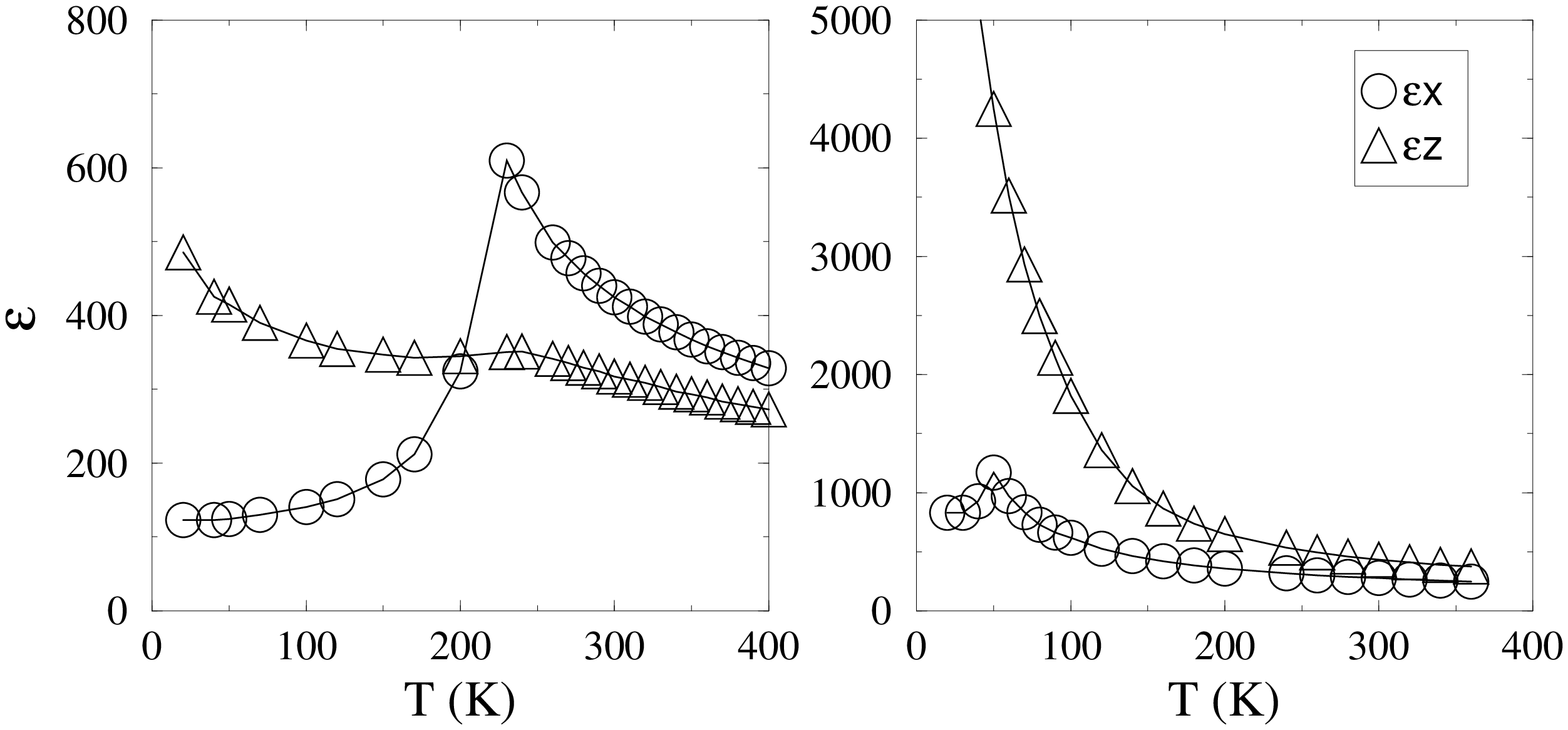}
\includegraphics[scale=0.64]{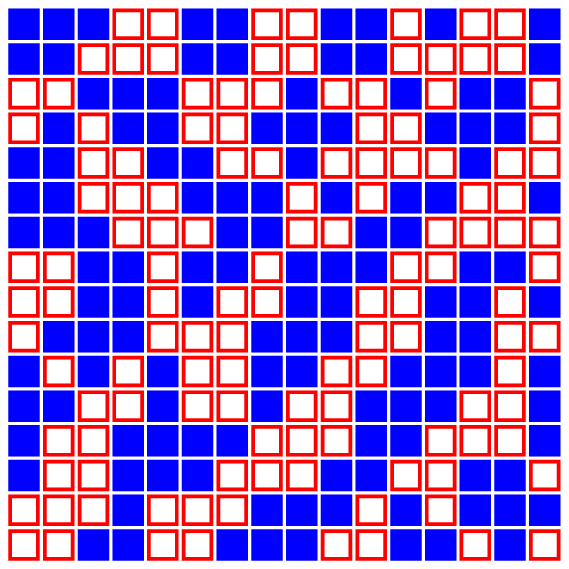}
\includegraphics[scale=0.25]{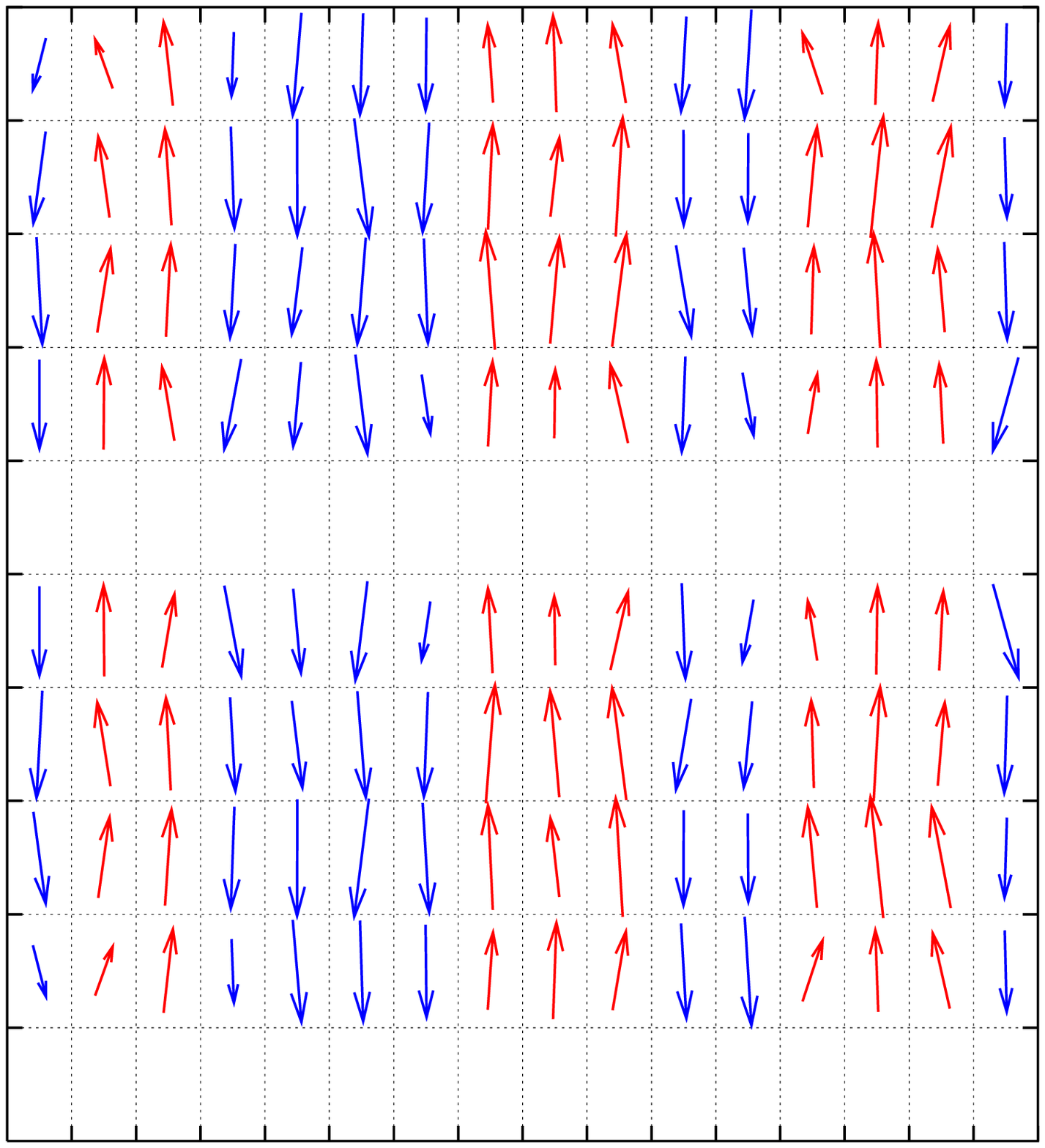}
\caption{Top: Dielectric constant as a function of temperature for (left)
films ($F$) and (right) epitaxial films $EF$ ($l$=4, $d$=1), sandwiched 
between ``bad" electrodes. 
Bottom: A horizontal slice on the left ($+z$ and $-z$ polarized sites 
are symbolized by empty and filled boxes respectively) and a vertical 
cross-section on the right for $16\times16\times4$ system for the same 
configuration at 110 K showing domain formation.}
\end{figure}

Ferroelectricity in the ultra-thin films of BaTiO$_3$ sandwiched between ``bad" electrodes
($d=1$) is qualitatively different from its bulk behavior. In this case, the depolarization field is
quite sizeable and effective in suppressing $P_z$. The dielectric constant (see Fig. 4, top panel)
$\epsilon_{zz}$ exhibits a divergence near $T=0$ K, but no signature of a phase
transition to a ferroelectric state with polarization $P_z$. The tetragonality
of the high temperature paraelectric phase is also apparent in the uniaxial symmetry
of dielectric constant. The transition to FE phase ordered with in-plane polarization
occurs at around 70$\pm$10 K for epitaxial thin films. For the films without epitaxial
constraint, this transition occurs at a higher temperature ($\sim$ 210 K).
To understand the nature of low temperature phase with very high $zz$ dielectric response,
we examine the snapshots of dipolar configurations of these films. They reveal (Fig. 4, bottom panel)
formation of vertical ((010) planes) stripe-like domains with polarization along $z-$axis which form
arbitrary patterns in the (001) planes. Our results are very similar to those observed
experimentally by Fong et al \cite{fong}. We find that these domains are stable throughout the
duration of our simulation (about 0.1 nano sec) and observed at temperatures as high as 
300 K. It is clear that these stripe-like domains are stabilized through minimization of
the energy cost associated with depolarization field. Formation of similar 
domain structure 
was also predicted by Levanyuk et al \cite{levanuk}.

Origin of most of our observations can be understood through Fourier analysis of the
dipolar configurations noting the mirror boundary conditions and vanishing of dipoles
in the gap region. For example, $P_z$ can be only an even function, where as
$P_x$ is always an odd function of $z$; hence $q=0$ component of $P_x$ has to
be zero (for the doubly periodic cell). This leads to fluctuations of the in-plane
electric dipoles whose length-scale is determined by the film thickness. Since these
dipoles couple with both the strain and out-of-plane dipoles, ferroelectricity
even in the absence of depolarization field exhibits thickness dependence.

In summary, we have shown that the ferroelectricity in thin films sandwiched between 
electrodes exhibits nano-thickness dependence even if the depolarization field vanishes.
Secondly, the presence of electrodes and their nature have wide-ranging
consequences to ferroelectricity in ultra-thin films from simple size dependence to
stabilization of phases with stripe-like domains. The origin of these phenomena
lies in interaction between local dipoles and their electrostatic images in electrodes.

UVW thanks P. Ayyub and R. Budhani for useful discussion; TN thanks JNCASR for local
hospitality. 
We thank International Frontier Center for Advanced Materials (IFCAM) 
of IMR who supported UVW to visit to Sendai and authors' collaborative study in
ferroelectrics
and the Centre for Computational Materials Science at JNCASR for financial support.


\begin{references}
\bibitem{scott} J. F. Scott, C. A. P. de Araujo, Science {\bf 246}, 1400 (1989)
\bibitem{Ahn} C. H. Ahn, K. M. Rabe, J. M. Triscone, Science {\bf{303}}, 488
(2004)
\bibitem{fong} D. D. Fong, G. B. Stephenson, S. K. Streiffer, J. A. Eastman,
O. Auciello, P. H. Fuoss, C. Thompson, Science, {\bf 304}, 1650 (2004)
\bibitem{choi} K. J. Choi, M. Biegalski, Y. L. Li, A. Sharan, J. Schubert,
R. Uecker, P. Reiche, Y. B. Chen, X. Q. Pan, V. Gopala, L. Q. Chen,
D. G. Schlom, C. B. Eom, Science {\bf 306}, 1005 (2004)
\bibitem{lines} M. E. Lines and A. M. Glass, {\emph{Principles and Applications
of Ferroelectrics and Related Materials}} (Oxford University Press, New York,
 1977)
\bibitem{ghosez} J. Junquera and P. Ghosez, Nature {\bf 422}, 506 (2003)
\bibitem{fu-cohen}H. X. Fu, R. E. Cohen, Nature, {\bf{403}}, 281 (2000)
\bibitem{pertsev} N. A. Pertsev, A. G. Zembilgotov, A. K. Tagantsev, Phys. Rev.
Lett. {\bf{80}}, 1988 (1998)
\bibitem{vand} O. Di\'{e}guez, S. Tinte, A. Antons, C. Bungaro, J. B. Neaton,
K. M. Rabe, D. Vanderbilt, Phys. Rev. B {\bf{69}}, 212101 (2004)
\bibitem{zhong} W. Zhong, D. Vanderbilt, K. M. Rabe, Phys. Rev. Lett. {\bf 73},
1861 (1994)
\bibitem{rabe} W. Zhong, D. Vanderbilt, K. M. Rabe, Phys. Rev. B
{\bf 52}, 6301 (1995)
\bibitem{umesh} U. V. Waghmare, K. M. Rabe, Phys. Rev. B {\bf 55}, 6161 (1997)
\bibitem{dawber} M. Dawber, P. Chandra, P. B. Littlewood, J. F. Scott, J. Phys.:
Condens. Matter, {\bf 15}, L393 (2003)
\bibitem{umesh1} U. V. Waghmare, E. J. Cockayne, B. P. Burton, Ferroelectrics
{\bf 291}, 187 (2003)
\bibitem{umesh2} B. P. Burton, E. J. Cockayne, U. V. Waghmare, Phys. Rev B
{\bf 72}, 064113 (2005)
\bibitem{nose} S. D. Bond, B. J. Leimkuhler, B. B. Laird, J. Comput. Phys.
{\bf 151}, 114 (1999)
\bibitem{SM}  K. Binder, in Phase Transitions and Critical Phenomena, Vol. 5B, edited by C. Domb, M.S. Green, Academic Press (1976)
\bibitem{leejun}L. Kim, J. Kim, U. V. Waghmare, D. Jung, J. Lee, 
Phys. Rev. B {\bf{72}}, 214121 (2005)
\bibitem{levanuk} A. M. Bratkovsky and A. P. Levanyuk, 
Phys. Rev. Lett. {\bf 84} 3177; ibid 86 3642 (2001).
\end{references}
\end{document}